\documentclass{aa}  
\usepackage{orcidlink}
\usepackage{graphicx}
\usepackage{txfonts}

\usepackage{xcolor}

\begin{document}

   \title{The Role of Preceding CMEs and SIRs in Enhancing Shock Acceleration of Electrons}
   \titlerunning{Preceding CMEs and SIRs in Electron Acceleration}

   \author{
       Xiaomin Chen\inst{1,2}~\orcidlink{0009-0003-3760-705X}
        \and Seve Nyberg\inst{2}~\orcidlink{0000-0003-2672-5491}
        \and Rami Vainio\inst{2}~\orcidlink{0000-0002-3298-2067}
        \and Immanuel Christopher Jebaraj\inst{2}~\orcidlink{0000-0002-0606-7172}
        \and Chuan Li\inst{1}~\orcidlink{0000-0001-7693-4908}
        \and Nina Dresing\inst{2}\thanks{Corresponding author: nina.dresing@utu.fi}~\orcidlink{0000-0003-3903-4649}
        }

   \institute{School of Astronomy and Space Science and Key Laboratory of Modern Astronomy and Astrophysics, Nanjing University, Nanjing 210023, People’s Republic of China
   \and Department of Physics and Astronomy, University of Turku, 20014 Turku, Finland
   }

   \date{Received}

   \abstract
{The role of large-scale pre-existing interplanetary structures, including preceding coronal mass ejections (CMEs) and stream interaction regions (SIRs), in shaping the shock acceleration environment for energetic electrons remains not fully understood. In this study, we investigate nine interplanetary shocks observed by the Solar Terrestrial Relations Observatory (STEREO) that are associated with significant MeV electron enhancements, as such enhancements are rarely observed at interplanetary shocks. We combine remote-sensing observations, drag-based modeling, and in-situ measurements to analyze the shock propagation through or into pre-existing interplanetary structures. Eight of the nine events are associated with a preceding slow or intermediate-speed CME, while six shocks are in-situ observed propagating within preceding ICMEs, indicating that large-scale upstream trapping structures are a common feature of these events. Further analysis identifies three distinct scenarios associated with enhanced electron acceleration: shocks propagating into preceding ICMEs, shock-SIR interactions, and direct injection of flare-accelerated electrons into SIRs. As a representative shock-in-ICME event, the 2012 January 29 low-$\beta$, quasi-perpendicular shock ($\theta_{Bn}\sim87^\circ$) is further investigated using observations together with one-dimensional Monte Carlo test-particle simulations of a shock propagating into a large-scale upstream magnetic loop. The simulation suggests that the upstream loop prolongs electron residence near the shock and substantially enhances acceleration efficiency. These results demonstrate that large-scale interplanetary structures can precondition the upstream magnetic environment, providing favorable conditions for prolonged electron residence and efficient shock acceleration.
   }

   \maketitle

\nolinenumbers
\section{Introduction}

Interplanetary shocks driven by fast coronal mass ejections (CMEs) are capable of accelerating energetic particles over a broad energy range. Statistical studies of interplanetary shocks near 1 au have established that these shocks can efficiently accelerate ions, with the acceleration efficiency depending strongly on the shock strength and geometry \citep[e.g.,][]{van1984energy,zank2007particle}. For electrons, recent Parker Solar Probe observations have revealed efficient acceleration at interplanetary shocks close to the Sun \citep[e.g.,][]{2024ApJ...968L...8J,2024ApJ...976L...7J,2025ApJ...987...31W}. At larger heliospheric distances, a survey of 475 shocks observed by the Solar Terrestrial Relations Observatory (STEREO) found shock-associated electron intensity enhancements at energies below 100 keV for only $\sim$1\% of the shocks, indicating that interplanetary shocks near 1 au are generally inefficient electron accelerators \citep{dresing2016efficiency}. Acceleration of electrons to relativistic speeds by shocks at 1 au is very rare. Among 587 interplanetary shocks observed by STEREO, only nine events were identified with statistically significant MeV electron enhancements, all associated with fast (>900 km s$^{-1}$) CMEs \citep{stereo_shock_acc}. The physical conditions that enable these rare MeV electron acceleration events are still unclear.

Shock drift acceleration (SDA) is an efficient mechanism for electron acceleration at quasi-perpendicular shocks \citep[e.g.,][]{sda_particle_trajectory,ball2001shock}, and its role has been supported by previous observations of electron acceleration at interplanetary shocks near 1 au \citep{shock_1au,2019ApJ...875..104Y}. In SDA, electrons gain energy through gradient-$B$ drift within the shock transition. As such, the details of the shock microstructure such as the magnetic overshoot, the scale of the ramp, and the separation between magnetic and electric field gradients influence the acceleration efficiency significantly. Stochastic shock drift acceleration (SSDA) extends this framework by incorporating pitch-angle scattering, allowing electrons to undergo multiple reflections within the shock transition and reach higher energies \citep[e.g.,][]{katou2019,amano2020}. Efficient electron acceleration often requires the demagnetization of the electron and its subsequent trapping within the shock transition layer by small scale fields, allowing particles to experience multiple encounters within the shock ramp \citep{2020ApJ...895...59G}. Magnetic trapping has therefore been considered as an important mechanism for extended electron residence near the shock, enabling repeated interactions with the shock and associated electromagnetic fluctuations and thereby enhancing electron energization \citep[e.g.,][]{shi2025compound}. Previous studies have proposed several trapping configurations, including traps involving converging magnetic mirrors \citep{gisler1990electron} and trapping between shock ripples \citep{xu2025electron}. However, these studies mainly involve trapping at the local shock structure and substructure. For ICME-driven interplanetary shocks, the upstream region may additionally contain large-scale interplanetary magnetic transients, such as preceding ICMEs. The influence of such upstream structures on electron acceleration has received little attention in in-situ shock observations.

Previous observations have demonstrated that large-scale interplanetary structures, including preceding CMEs and stream interaction regions (SIRs), can substantially modify solar energetic particle (SEP) transport by altering magnetic connectivity and facilitating particle confinement, which may further contribute to additional acceleration \citep[e.g.,][]{dresing2016efficiency,dresing2018long,guo2018modeling,palmerio2021cme,Lario_SIR,2024ApJ...967L..33C}. Furthermore, enhanced energetic particle intensities have been observed within complex ICME structures formed by CME–CME interactions, indicating local particle acceleration \citep[e.g.,][]{2008SoPh..252..409S,2019ApJ...885...54X,chen2025local}. Recent numerical simulation further suggests that non-uniform background solar wind structures can distort CME-driven shocks, leading to spatial variations in shock compression, obliquity, and particle injection efficiency \citep{Nicolas_SIR}. Simulation of interacting CMEs has demonstrated that multiple acceleration processes, including the macroscopic collapsing magnetic trap (MCMT) effect, can significantly enhance energetic particle acceleration \citep{2024ApJ...967L..35N}. Despite these findings, direct in-situ observations of shocks propagating into preceding ICMEs remain important for determining how these structures influence particle acceleration. 

In this study, we investigate nine interplanetary shocks observed by STEREO that were identified by \cite{stereo_shock_acc} as being associated with significant MeV electron enhancements, all linked to fast CMEs (>900 km/s) \footnote{One event listed as a CME-driven shock is reinterpreted here as an SIR-associated shock (Sec. \ref{sec:flare_contribution}).} and, in some cases, preceding shocks. Section \ref{sec:instrument} briefly introduces the observational data we used in this study. In Sec. \ref{sec:source_propagation}, we identify the solar source of each event and reconstruct the interplanetary propagation of CMEs using a drag-based model (DBM). In Sec. \ref{sec:shock_preconditioned}, we investigate shock propagation through preconditioned solar wind and examine how preceding CMEs and SIRs influence the upstream environment and electron acceleration. Section \ref{sec:flare_contribution} presents a case study of flare accelerated particles injected into a SIR. In Sec. \ref{sec:simulation}, we simulate the 2012 January 29 event to explore electron acceleration in a shock-in-ICME configuration. Discussion and conclusions are presented in Sec. \ref{sec:conclusion}.  

\section{Observational Data}\label{sec:instrument}

We primarily analyze observations from the in-situ Measurements of Particles And CME Transients (IMPACT, \citealt{2008SSRv..136..117L}) and the Plasma and Suprathermal Ion Composition (PLASTIC, \citealt{2008SSRv..136..437G}) investigations aboard the STEREO-A (STA) and STEREO-B (STB) spacecraft. The datasets include magnetic field measurements from the Magnetometer (MAG), proton plasma parameters from PLASTIC, and electron pitch-angle distributions (PADs) from the Solar Wind Electron Analyzer (SWEA, \citealt{2008SSRv..136..227S}). Magnetic field observations are available at the nominal cadence of 8 s$^{-1}$, while burst-mode data (32 s$^{-1}$) are used to resolve fine shock structures. Proton plasma parameters are provided at 1 min cadence, and SWEA measures electron PADs every 30 s over energies from a few eV to several keV. Energetic electron observations are obtained from the High Energy Telescope (HET, \citealt{2008SSRv..136..391V}), using the three electron channels covering 0.7--1.4, 1.4--2.8, and 2.8--4.0 MeV. In Sec.~\ref{sec:simulation}, we additionally use the lowest-energy electron channel of SEPT ($\sim$60 keV) to extend the energy coverage. Radio dynamic spectra are obtained from STEREO/WAVES (SWAVES, \citealt{2008SSRv..136..487B}). Solar source regions are investigated using extreme-ultraviolet (EUV) and white-light observations from Sun Earth Connection Coronal and Heliospheric Investigation (SECCHI, \citealt[]{2008SSRv..136...67H}), together with Atmospheric Imaging Assembly (AIA, \citealt[]{2012SoPh..275...17L}) aboard Solar Dynamics Observatory (SDO) and Large Angle and Spectroscopic Coronagraph (LASCO, \citealt[]{brueckner1995large}) aboard Solar and Heliospheric Observatory (SOHO).

\section{Solar Source and Interplanetary Evolution}\label{sec:source_propagation}
\subsection{Solar Source Identification of ICMEs}

\begin{figure*}[htb!]
    \begin{center}
    \includegraphics[width=0.9\textwidth]{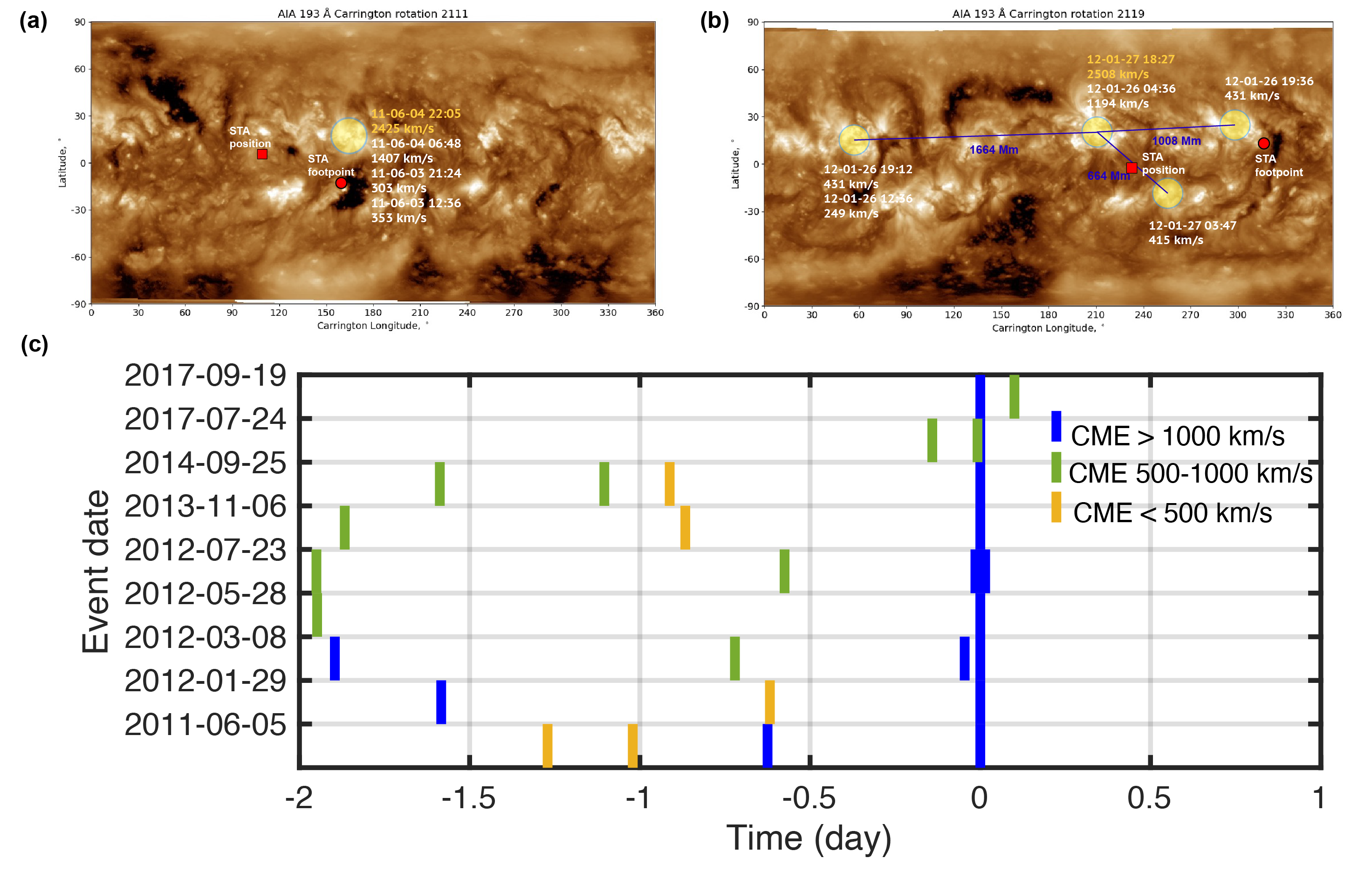}
    \caption{Source regions of the major and preceding CMEs in two events and temporal sequences of CMEs for all nine ESP events. Panels (a) and (b) present the source regions of the major and preceding CMEs in the 2011 June 5 and 2012 January 29 events, respectively, marked by yellow circles. The onset time of the major CME is highlighted in yellow. The blue lines indicate the great-circle distances between the two active regions on the solar surface, while the red dot and square mark the footpoint and the line-of-site position of STA, respectively. Panel (c) shows the timing relationship between the major CMEs and their preceding CMEs for all events. \label{fig:source}}
    \end{center}
\end{figure*}

Using LASCO CME catalog \footnote{\url{https://cdaw.gsfc.nasa.gov/CME_list/}}, we identified the major CME and preceding CMEs that occurred within the previous two days, obtaining their onset times and projected speeds from the catalog. Next, we examined the EUV and white-light observations from STEREO and near-earth spacecraft to locate the associated eruptive active region. In white-light observations, we required a clearly identifiable bright leading edge in LASCO/C2 images. CMEs labeled as poor or very poor in the LASCO catalog were generally not included, as their source regions are difficult to identify reliably. A few events that appeared weak in C3 but exhibited clear signatures in C2 were still included in our analysis. Consequently, our selection may omit some stealth CMEs or very weak events. In addition, some major events correspond to twin-CME eruptions (e.g., the 2012 July 22 event), which are treated as a single CME in the catalog. As shown in Fig. \ref{fig:source}, the CME onset times and speeds are labeled at their corresponding source locations on 193\AA~synoptic maps from SDO/AIA. The synoptic maps are obtained from \cite{synoptic_map}. Fig. \ref{fig:source1} in Appendix \ref{sec:sources} provides the same information for all nine events.

Since our focus is on electron acceleration by CME-driven shocks, we evaluated the potential interaction between major and preceding CMEs on the basis of the possible longitudinal overlap of their shock fronts. As the major CMEs in our sample are generally wide and fast, we adopt a relatively broad criterion and include any preceding CME whose source region lies within 1000~Mm of that of the major CME. The separation distance is estimated by multiplying the angular distance between the two active regions by the solar radius. Panel (c) of Fig. \ref{fig:source} presents the temporal sequence of the preceding and major CMEs. We classify CMEs as fast (> 1000 km/s), intermediate-speed (500--1000 km/s), and slow (<500 km/s). In one third of the events, a fast preceding CME occurs within one day before the major eruption, likely leading to CME interactions. In eight out of nine events, the major CME is preceded by at least one slow or intermediate-speed CME. Although such preceding CMEs are not expected to be efficient particle accelerators on their own, their potential role in modifying the ambient solar wind and influencing the acceleration efficiency of the subsequent fast CME is also a key focus of this study. One event has no preceding CME, but is followed by an intermediate-speed CME approximately 2 h later. 

To illustrate the relative positions of the solar sources and the spacecraft, we marked the spacecraft line-of-site projected positions with squares in Fig. \ref{fig:source} and Fig. \ref{fig:source1}. In four out of nine events, the spacecraft position is within 30$^\circ$ from the eruption site, suggesting a shock nose geometry. The remaining five events corresponded to flank encounters, including two eastern-flank and three western-flank configurations. We also estimated the nominal magnetic connectivity using the Solar-MACH tool \citep{solar_mach}, which are marked by dots in Fig. \ref{fig:source} and Fig. \ref{fig:source1}. In roughly two-thirds of the events, the spacecraft footpoints were located west of the eruption site. The actual magnetic connectivity may differ because preceding CMEs can substantially modify the interplanetary magnetic field.

\subsection{Interplanetary Propagation and Inferred Interaction}\label{sec:propagation}

To investigate CME propagation and possible interactions from the source region to 1 au, we apply the one-dimensional DBM \citep{dbm_initial,dbm} to all major and preceding CMEs for each event. The DBM assumes that beyond a certain distance the Lorentz force becomes negligible, and CME propagation is primarily governed by the interaction with the ambient solar wind. The drag acceleration is given by $a = -\gamma (v-w)|v-w|$, where $a$ and $v$ are the CME acceleration and speed, $w$ is the ambient solar wind speed, and $\gamma$ is the drag parameter. 

\begin{figure}[htb!]
    \begin{center}
    \includegraphics[width=1\linewidth]{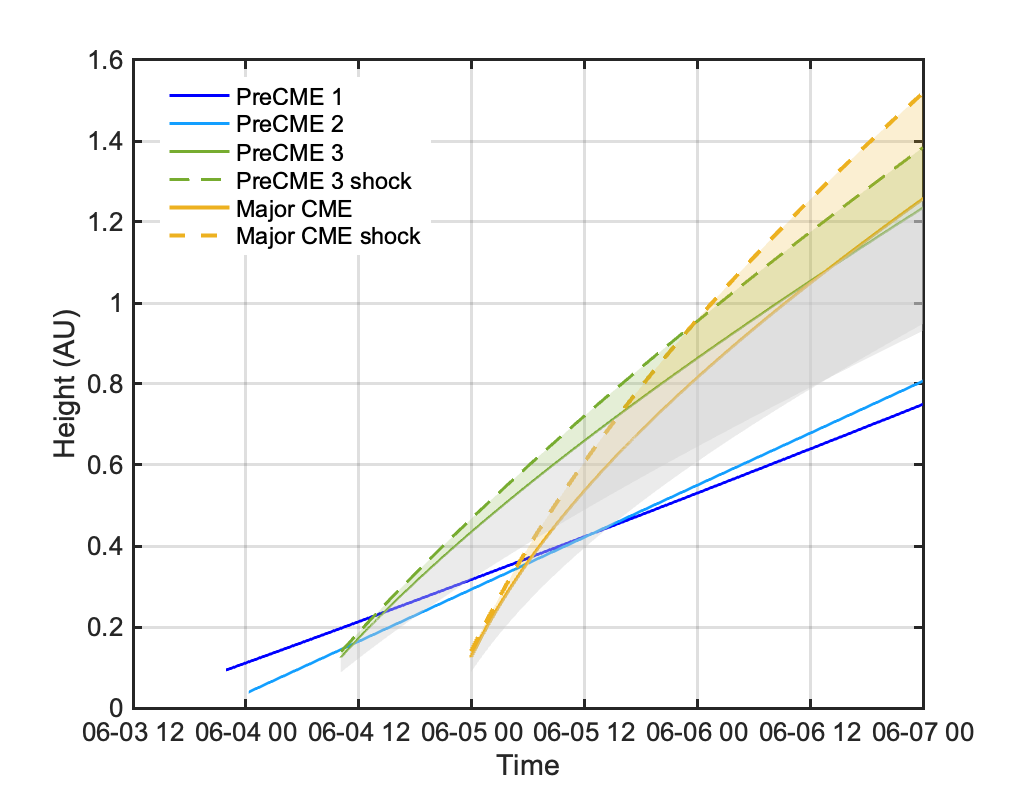}
    \caption{Height–time profiles derived from the DBM for the 2011 June 5 event. The green shaded region denotes the sheath of PreCME 3, while the yellow shaded region marks the sheath of the major CME. The gray shaded areas indicate the radial widths of these CMEs. \label{fig:dbm}}
    \end{center}
\end{figure}

The choice of $\gamma$ and $w$ has been extensively investigated by previous studies. For our statistical sample of nine events, following the parameter combinations explored in \cite{dbm_initial1} and \cite{DBM_param}, we adopt common used values of $\gamma = 0.2\times10^{-7} $ km$^{-1}$ and $w = 450$ km/s for modeling CME propagation. Since our focus is on the particle acceleration by CME-driven shocks, we additionally model the shock propagation separately. Previous studies have shown that CME-driven shocks undergo weaker de-acceleration than the CME ejecta and are therefore represented with a lower $\gamma$ value \citep{shock_dbm,DBM_param}. Accordingly, we set $\gamma = 0.1\times10^{-7} $ km$^{-1}$ and $w = 450$ km/s for the shock DBM. The initial conditions are taken as the second-order projected speed at the final height in the LASCO CME catalog, together with the corresponding time and height. Considering that below $\sim$30 $R_\odot$, the speeds of the CME front and its driven shock is comparable, we use the same initial speed and time for the CME and the shock, and assume an initial sheath thickness of 3 $R_\odot$. To estimate the radial width of the CME, we adopt the radial expansion fit derived from ICME observations over 0.3--5.4 au by \cite{radial_thickness}, expressed as $S(R)=(0.25\pm0.01)R^{0.92\pm0.07}$ (au). $S(R)$ is the radial width and $R$ is the heliospheric distance of the CME. This prescription is primarily applied to fast and intermediate-speed CMEs, as most of the slow CMEs identified in our sample are caught up by subsequent CMEs within 0.3 au. In Fig. \ref{fig:dbm}, the lower boundary of the gray shaded region represents the inferred trailing edge of each CME. 

As an example, we derived the height-time profiles using the DBM for the 2011 June 5 event, which corresponds to Panel (a) of Fig. \ref{fig:source}. As shown in Fig. \ref{fig:dbm}, the trajectory of preceding CME 3 (hereafter PreCME 3) intersects those of preCME 1 and preCME 2 within 0.3 au, suggesting possible interaction or merging. The major CME-driven shock is expected to intersect the trailing edge of preCME 3 at $\sim$0.4 au and subsequently propagate into its ejecta, marking the onset of a direct CME--CME interaction. The shock is likely to propagate through preCME 3 and merge with its driven shock within 1 au. We note that the DBM here assumes constant values of the drag parameter $\gamma$ and the ambient solar wind speed $w$, whereas both quantities are expected to vary with time and heliocentric distance. In particular, the preceding CME may reduce the ambient solar wind density, resulting in a smaller effective drag parameter for subsequent CMEs \citep{shock_dbm}. Therefore, the model does not explicitly account for CME–CME interactions or the temporal evolution of the background solar wind, but instead assumes that each CME propagates independently in a uniform ambient medium. The results should therefore be regarded as a simplified estimate of possible trajectory crossings, rather than a realistic treatment of CME–CME interactions.

\section{Shock Evolution in a Preconditioned Solar Wind}\label{sec:shock_preconditioned}

\subsection{Shock Propagation inside a Preceding CME}\label{sec:shock_in_cme}

We investigate a typical process in CME–CME interactions, namely the major CME-driven shock propagating through or into a preceding CME. Focusing on the shocks associated with the major CMEs on 2012 January 29 and 2012 May 28, we combine remote-sensing and in-situ observations to investigate their upstream environments, shock structures, and associated particle acceleration.

The parameters of the major shocks in the two events are summarized in Tab. \ref{tab:shock}. Following the sliding averaging window method described in \cite{shock_parameter}, we derived the shock speed $V_{sh}$, shock normal angle $\theta_{Bn}$, magnetic field ratio $B_d/B_u$, fast magnetosonic Mach number $M_{fms}$, upstream proton beta $\beta$, upstream Alfv\'en speed $V_A$, Alfv\'enic Mach number $M_A$. We use upstream and downstream intervals defined relative to the shock arrival: 2–10 min before the shock for upstream and a similar interval after the shock for downstream, with the minimum window spanning 3 min. The shock normal was determined using the MX3 method \citep{MX3}. Since the 1-min proton plasma data yield few upstream/downstream samples, the uncertainties are approximate estimates. The 29 January event represents an almost ideal perpendicular configuration with $\theta_{Bn} \sim 87^\circ$, together with a low upstream proton plasma $\beta$ of 0.08, consistent with a typical ICME environment. The 28 May shock is more oblique ($\theta_{Bn} \sim 68^\circ$)  with a moderate $\beta$ of 0.62. Notably, for the shock on 2012 May 28, the upstream and downstream region are highly variable, and the downstream averaging interval overlaps with an enhanced magnetic field structure, resulting in the large $B_d/B_u$. The local magnetic field enhancement near the shock is further examined in the 10-s zoom-in below.

\begin{figure*}[tp]
\centering
\includegraphics[width=0.9\linewidth,
    height=0.75\textheight,
    keepaspectratio]{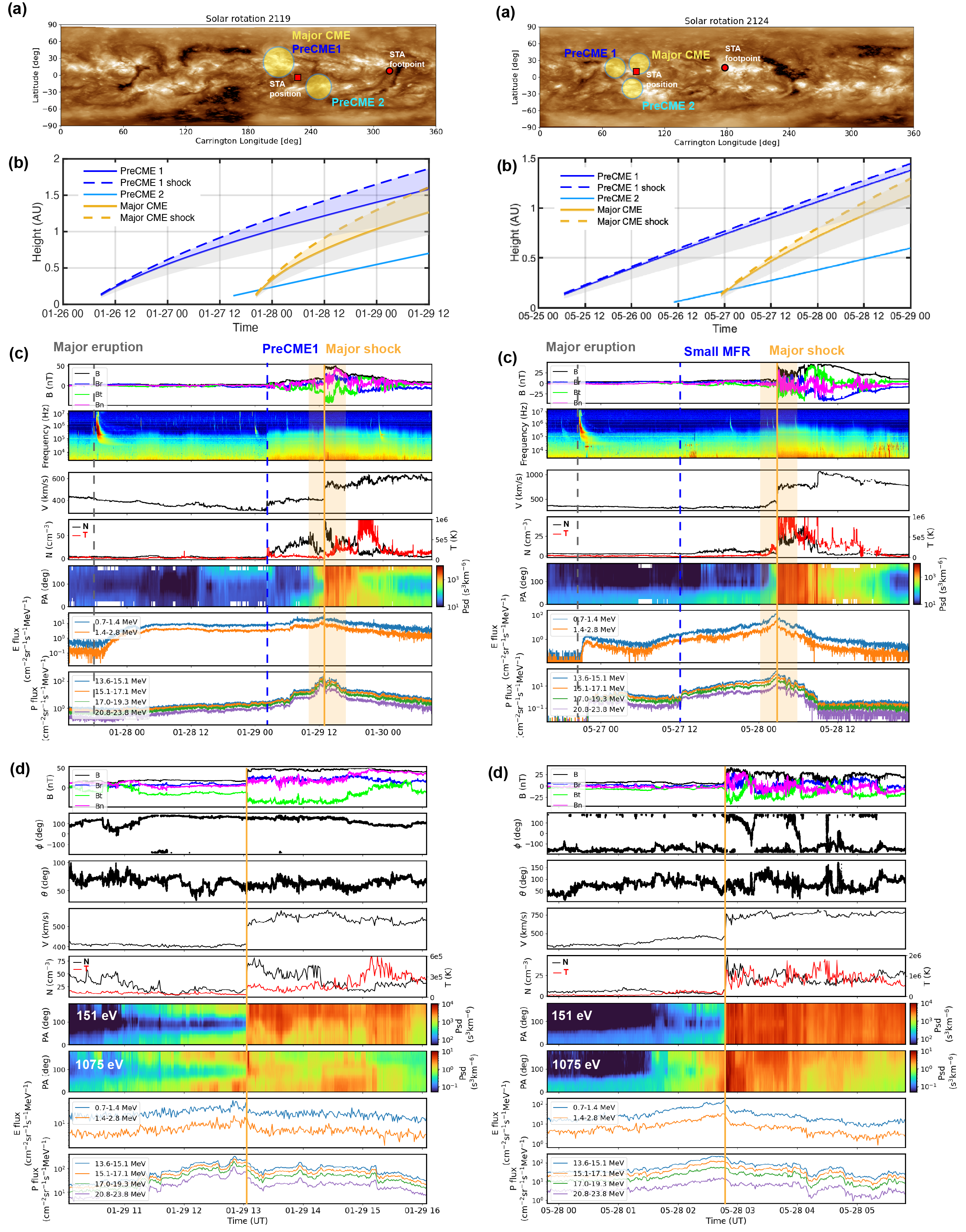}
\caption{\label{fig:shock_mfr}Two shock-in-PreCME events: 2012 January 29 (left) and 2012 May 28 (right). (a) Synoptic maps showing the major and preceding CMEs. (b) DBM propagation results. (c) in-situ observations from the major eruption to the shock arrival, including the magnetic field, radio spectrum, solar wind plasma parameters, suprathermal ($\sim$151 eV) electron PADs, and HET energetic particle fluxes. The blue dashed line marks the arrival of the PreCME 1-driven shock (left) and the start of the preceding small flux rope (right), while the orange line indicates the arrival of the major shock. The yellow shaded region denotes the $\pm$3h interval around the shock. (d) Zoom-in view ($\pm$3 h) around the shock, shown in the same format as Panel (c) but without the radio spectrum; electron PADs at 151 eV and 1075 eV are displayed. }
\end{figure*}

The left panels (a)--(d) in Fig. \ref{fig:shock_mfr} present the shock event on 2012 January 29. It is associated with a major CME erupted on January 27 at 18:27 UT with a linear speed of 2508 km/s in the LASCO catalog. Two preceding CMEs occurred earlier: preCME 1 on January 26 at 04:36 UT with a speed of 1194 km/s and preCME 2 on January 27 at 03:47 UT with a speed of 415 km/s. As shown in Panel (b), the DBM results indicate that the major CME driven shock may catch up with the PreCME 2 within 0.3 au. It is further expected to reach the trailing edge of PreCME 1 at $\sim$0.9 au, but not to reach its leading edge within 1 au. Thus, there is no clear indication of shock merging in this event. Panel (c) presents the in-situ observation by STA. A weak CME-driven shock arrived at STA at 02:13 UT on January 29, which we attribute to the shock driven by PreCME 1. The major CME-driven shock is observed at 13:04 UT, about 13 h later than the DBM prediction, potentially reflecting de-acceleration due to the preceding eruptions. The inferred radial width of PreCME 1, estimated from its duration multiplied by the average solar wind speed, is $\sim$0.16 au, smaller than the typical mean value of $\sim$0.25 au at 1 au \citep{radial_thickness}. This reduced width likely results from both the compression of PreCME 1 by the propagating major CME-driven shock and the partial sampling of the compressed ejecta by STA. Similar compression effects have been reported by previous studies of CME-CME interactions \citep[e.g.,][]{wang2003interplanetary,xiong2006magnetohydrodynamic,lugaz2015extreme,shen2018shock,chi2024direct}. In addition, PreCME 1 does not exhibit a single coherent flux rope but instead contains multiple embeded flux-rope-like substructures, with the one immediately upstream of the shock examined in detail below. The MeV electron intensity began to increase shortly after the major eruption and exhibited a second enhancement within PreCME 1 before reaching the peak near the major CME-driven shock.     

\begin{figure}[htb!]
    \begin{center}
    \includegraphics[width=1\linewidth]{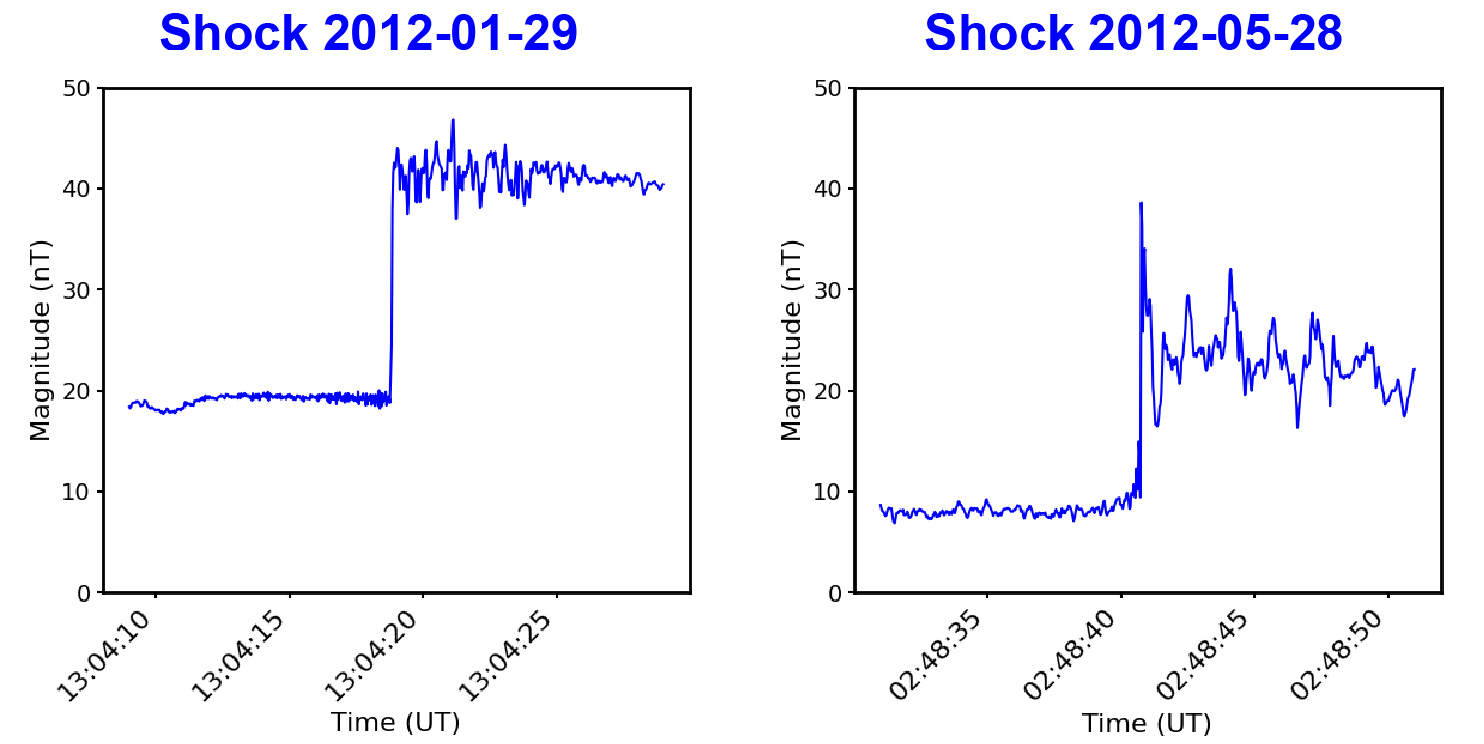}
    \caption{Magnetic field magnitude within $\pm$10 s of the shock crossing for the two events shown in Fig. \ref{fig:shock_mfr}. \label{fig:overshoot}}
    \end{center}
\end{figure}

The left panel (d) of Fig. \ref{fig:shock_mfr} shows a zoom-in view ($\pm$3 hr) of the shock on January 29. The shock discontinuity is very sharp, while the upstream and downstream magnetic fields remain relatively stable. In the 2 h upstream of the shock, a sub-flux rope within PreCME 1 is observed, accompanied by a gradual rotation of the magnetic field azimuthal angle ($\phi_B$). The smooth variation of $\phi_B$ across the shock indicates that the shock does not induce significant magnetic field rotation. The solar wind temperature, velocity, and density all exhibit clear jumps at the shock. Electrons at 151 eV and 1075 eV show well-defined bi-directional streaming upstream for about 2 h, indicating a stable closed magnetic structure. Using the observed duration ($\sim$2 h) and an average solar wind speed of $\sim400$ km/s, we estimate that the spacecraft traversed the closed magnetic structure spanning $\sim2.9\times10^6$ km ($\sim0.02$ au, or $\sim4.1$ $R_\odot$). MeV electrons gradually increase in the upstream region, reaching a peak roughly 10 min before shock arrival and exhibiting a second distinct spike at the shock. As shown in Tab. \ref{tab:shock}, this is a typical quasi-perpendicular shock. The left panel in Fig. \ref{fig:overshoot} shows a zoom-in of the magnetic field magnitude within $\pm$10 s of the shock crossing, revealing a sharp shock ramp with a duration of only $\sim$0.3 s. The magnetic field increases from $\sim$ 18 nT upstream to $\sim$43 nT downstream, consistent with a clean and well-defined transition. A brief wave analysis is presented in Fig.~\ref{fig:wave}. The upstream region is generally quiet, except for high-frequency ($\sim$10 Hz) fluctuations observed during the $\sim$8 s immediately ahead of the shock. Their possible role is further discussed in Sec.~\ref{sec:simulation} and Sec.~\ref{sec:conclusion}. The well-defined quasi-perpendicular shock suggests favorable conditions for SDA, which predicts energy gain during a single shock encounter. However, the presence of a large-scale upstream magnetic flux rope may allow shock-reflected electrons to remain near the shock and experience repeated encounters, potentially leading to much higher energies. This possibility is investigated using numerical simulations in Sec. \ref{sec:simulation}.

The right panels (a)--(d) in Fig. \ref{fig:shock_mfr} present the shock event on 2012 May 28. The associated major CME erupted on 2012 May 26 at 20:57 UT, with a linear speed of 1966 km/s. Two preceding CMEs occurred earlier: PreCME 1 on May 24 at 22:12 UT with a speed of 563 km/s and PreCME 2 on May 26 at 04:48 UT with a speed of 276 km/s. Both PreCMEs are relatively slow and narrow, implying a weaker modification of the upstream solar wind compared to the previous case. As shown in panel (b), the DBM results suggest that the major CME may catch up with PreCME 2 within 0.3 au, reach the trailing edge of PreCME 1 at $\sim$0.8 au, but not its leading edge within 1 au. These results suggest CME interactions without clear evidence of shock merging. Panel (c) shows the in-situ observations by STA. A weak and long-lasting magnetic flux rope is observed for about 14 h upstream of the major shock, with the major shock propagating into this structure. Although its source cannot be clearly identified, its timing is consistent with the DBM predicted arrival of PreCME 1. But it may also be related to another weak CME that was not clearly detected in coronagraph observations. The major shock reached STA at 02:48 UT on May 28, earlier than predicted by the DBM. The sheath exhibits a complex magnetic field structure with multiple sub-structures. A discontinuity near the major ejecta is observed at $\sim$ 09:00 UT, characterized by increases in the solar wind speed and magnetic field magnitude and decreases in proton temperature and density. These signatures are inconsistent with a typical shock and may instead reflect CME interactions. Following the major eruption at the Sun, the MeV electron intensity rapidly reached an initial peak and then gradually declined. It subsequently rose again before the spacecraft entered the upstream magnetic structure and reached peak at the shock. The MeV proton intensity exhibited a clear enhancement within the structure before the arrival of the major shock. 

As shown in the right panel (d) of Fig.~\ref{fig:shock_mfr}, the May 28 event exhibits similar upstream signatures of suprathermal electrons, although the evidence for trapping is less pronounced. The 151 eV electrons exhibit a weaker and more diffuse bi-directional streaming that lasts for about 1 h. At higher energy (1075 eV), the PAD becomes more isotropic, indicating stronger pitch-angle scattering near the shock region. The weaker and more diffuse bi-directional streaming suggests that electrons remain partially confined upstream of the shock, but less effectively than in the January 29 event. In the 10-s zoom-in observaiton in Fig. \ref{fig:overshoot}, the shock ramp displays a pronounced overshoot, with the magnetic field magnitude reaching $\sim$ 42 nT, or $B_{max}/B_{up}$ $\sim$ 4.8 relative to the upstream 10-s average of $\sim$ 8.8 nT. Following the overshoot, the magnetic field is relatively steady at $B_d$ $\sim$ 23 nT, corresponding to $B_d/B_u\sim$2.6. Such overshoot could enhance magnetic mirroring within the shock transition \citep[e.g.,][]{giacalone1991effect}. The MeV electron flux gradually increases in the upstream region, reaches a peak very close to the shock arrival time, followed by a drop closely after the shock and then a slow decay. This temporal profile suggests that the acceleration occurs primarily near the shock front. 

These results suggest that both the large-scale upstream magnetic topology and the local shock structure play important roles in electron acceleration. The January 29 event exhibits stronger upstream magnetic confinement, whereas the May 28 event is characterized by a more structured shock transition that may enhance stochastic acceleration. Together, these observations indicate that efficient electron acceleration in shock-in-ICME events results from the combined effects of upstream confinement and local shock properties.

\begin{table*}
\caption{Parameters of the major shocks on 2012 January 29 and 2012 May 28 in Fig. \ref{fig:shock_mfr}, derived using the sliding averaging window method. The uncertainties represent the standard deviations obtained by varying the averaging window. $^{*}$ The large $B_{\rm d}/B_{\rm u}$ results from the enhanced magnetic field sampled by the downstream averaging interval.}
\label{tab:shock}
\centering
\renewcommand{\arraystretch}{1.3} 
\setlength{\tabcolsep}{6pt} 
\begin{tabular}{lccccccc}
\hline\hline
Date & $V_{sh}$ (km/s) & $\theta_{\mathrm{Bn}}$ ($^\circ$) & $B_d/B_u$ & $M_{fms}$ & $\beta$ & $V_A$ (km/s)  & $M_A$ \\
\hline
2012-01-29 13:04 & 505.3 $\pm$ 7.6 & 86.4 $\pm$ 0.5 & 2.7 $\pm$ 0.1 & 1.99 $\pm$ 0.05 & 0.08 $\pm$ 0.002 & 84.8 $\pm$ 2.0 & 2.05 $\pm$ 0.05  \\
2012-05-28 02:48 & 651.8 $\pm$ 7.9 & 68.4 $\pm$ 4.1 & 6.1 $\pm$ 0.4$^{*}$ & 3.00 $\pm$ 0.06 & 0.74 $\pm$ 0.12 & 66.1 $\pm$ 3.6 & 3.70 $\pm$ 0.16  \\
\hline
\end{tabular}
\end{table*}

\subsection{Shock Evolution in SIR Conditioned CME Interactions}\label{sec:SIR}

\begin{figure*}[tp]
\centering
\includegraphics[width=1\linewidth]{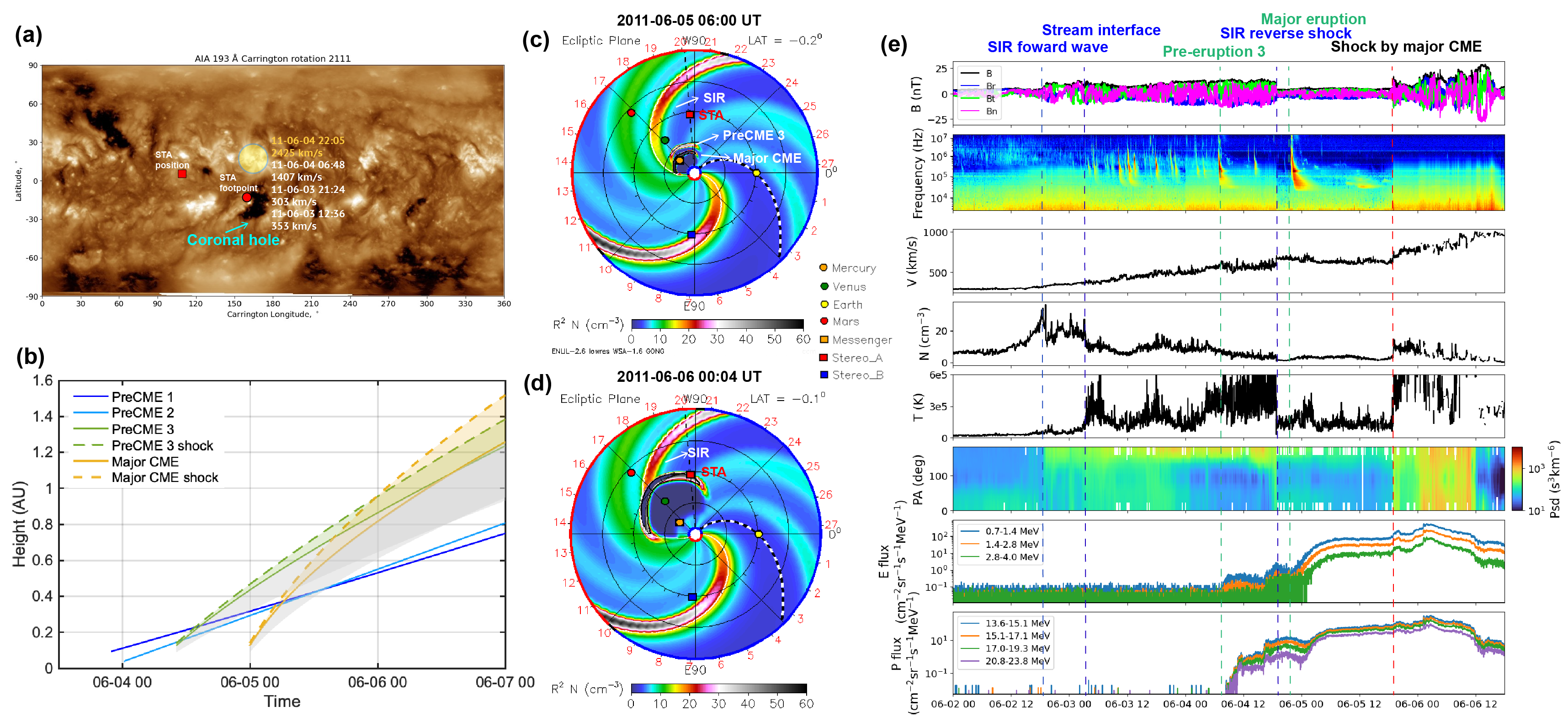}
\caption{\label{fig:cme_cir} Shock event on 2011 June 5. (a) Synoptic map showing the source regions of the major CME (yellow) and preceding CMEs (white). (b) DBM propagation results. (c)–(d) Snapshots from the WSA–ENLIL simulations. (e) in-situ observations from STA, including the magnetic field, radio dynamic spectrum, solar wind speed, proton density, temperature, PADs of 151 eV electrons, and temporal profiles of MeV electrons and protons.}
\end{figure*}

As shown in Fig. \ref{fig:cme_cir}, we next analyze a SIR-conditioned CME interaction event on 2011 June 5. In Panel (a), the major CME and three preceding CMEs originate from the same active region. A coronal hole is located to the south and magnetically connected with STA. Panel (b) shows the DBM results for this event, discussed in Section \ref{sec:propagation}. The trajectory of PreCME 3 intersects those of PreCME 1 and PreCME 2 within 0.3 au, suggesting possible interaction or merging. Taking the  radial width into account, the major CME-driven shock may enter the ejecta of PreCME 3 at $\sim$0.4 au, and may subsequently merge with its driven shock within 1 au. Meanwhile, the proximity of the CMEs to the coronal hole-associated high-speed stream (HSS) suggests that interaction with the HSS may begin at an early stage.

To further investigate the complex interaction scenario, Panel (c) and (d) show the Wang–Sheeley–Arge (WSA)–ENLIL +Cone simulation from the DONKI database\footnote{\url{https://kauai.ccmc.gsfc.nasa.gov/DONKI/view/WSA-ENLIL/2991/1}}. In this model, the background solar wind is generated by the WSA model \citep{wsa}, which provides inner boundary conditions at 21.5 $R_\odot$ for the ENLIL simulation \citep{enlil}, while CMEs are inserted as a Cone model \citep{cone} with prescribed kinematic and geometric parameters. In this simulation, only PreCME 3 and the major CME are included. The result reveals a well-defined SIR structure interacting with both CMEs. As the CMEs propagate, PreCME 3 and the major CME are predicted to merge at $\sim$0.6 au. By 1 au, the SIR appears to merge with the merged ejecta, indicating a complex interaction involving both CME–CME and CME–SIR interactions.

Panel (e) presents the in-situ observations by STA. As indicated by three blue dashed lines, a clear SIR is identified, which includes a forward wave, a stream interface, and a reverse shock. The forward wave is associated with enhancements in magnetic field and density, along with gradual increases in speed and temperature. Within the SIR compression region, the stream interface divides the slow, high-density, low-temperature solar wind from the fast, low-density, high-temperature component. At the reverse shock, we observe a sharp speed increase accompanied by decreases in magnetic field, temperature, and density.

At the time of the type III radio burst marking the solar eruption that produced PreCME 3, STA was located within the SIR, and a clear particle enhancement was observed inside the SIR. Local electron acceleration is detected at the reverse shock, suggesting that energetic electrons produced by the eruption and confined within the SIR compression region may provide seed particles for further acceleration at the shock. Between the reverse shock and the shock driven by the major CME (indicated by the red dashed line), electrons at 151 eV exhibit weak bi-directional streaming, likely due to reflection at the distant SIR reverse shock, which can produce a sunward-directed flux. When the merged CME-driven shock reaches STA, a pronounced electron flux enhancement is observed, with multiple intensity peaks in the sheath. The sheath itself exhibits a highly structured magnetic field, characterized by several large-angle magnetic field rotations, potentially associated with the presence of multiple preceding CMEs embedded within the sheath. Similar ICME-in-sheath structures, where preceding ICMEs are compressed and embedded within the sheath of subsequent CMEs, have been reported previously and associated with enhanced energetic particle populations in recent observations \citep{liu2020characteristics,chen2025local}.

Table \ref{tab:shock_sir} summarizes the parameters of the reverse shock and the shock driven by the merged CME. Both shocks are oblique, with moderate compression ratios, Mach numbers, and plasma $\beta$. In this event, the SIR acts as a large-scale compression region that can efficiently trap particles. In addition, during the eruption of PreCME 3, STA was still located inside the SIR, allowing flare-accelerated particles to be injected into the SIR and undergo further acceleration at the SIR shock pair. Owing to its radial extent from near the Sun to 1 au, the SIR can continuously interact with CMEs, potentially deflecting their propagation and modifying the upstream conditions of shocks, thereby influencing particle acceleration and transport. Numerical simulations have shown that interactions with corotating interaction regions can enhance CME deflection compared with propagation in a background solar wind environment \citep{liu2019numerical}. The MeV electron enhancement observed at the merged major CME-driven shock suggests that its acceleration efficiency may have been enhanced by earlier interactions with PreCME 3 and the SIR, although the shock evolution cannot be directly traced from the available observations.

\begin{table*}
\caption{Parameters of the reverse shock on 2011 June 4 and major shock on 2011 June 5 in Fig. \ref{fig:cme_cir}.}
\label{tab:shock_sir}
\centering
\renewcommand{\arraystretch}{1.3} 
\setlength{\tabcolsep}{6pt} 
\begin{tabular}{lccccccc}
\hline\hline
Date & $V_{sh}$ (km/s) & $\theta_{\mathrm{Bn}}$ ($^\circ$) & $B_d/B_u$ & $M_{fms}$ & $\beta$ & $V_A$ (km/s)  & $M_A$ \\
\hline
2011-06-04 18:48 & 365.2 $\pm$ 5.2 & 48.0 $\pm$ 2.6 & 2.6 $\pm$ 0.03 & 2.09 $\pm$ 0.03 & 0.30 $\pm$ 0.01 & 74.7 $\pm$ 5.3 & 2.24 $\pm$ 0.04  \\
2011-06-05 18:59 & 570.2 $\pm$ 11.0 & 50.9 $\pm$ 6.2 & 2.3 $\pm$ 0.1 & 2.53 $\pm$ 0.07 & 0.67 $\pm$ 0.13 & 64.6 $\pm$ 2.6 & 2.96 $\pm$ 0.09  \\
\hline
\end{tabular}
\end{table*}

\section{Flare Contribution to Seed Particles for Interplanetary Shock Acceleration} \label{sec:flare_contribution}

\begin{figure*}[tp]
\centering
\includegraphics[width=1\linewidth]{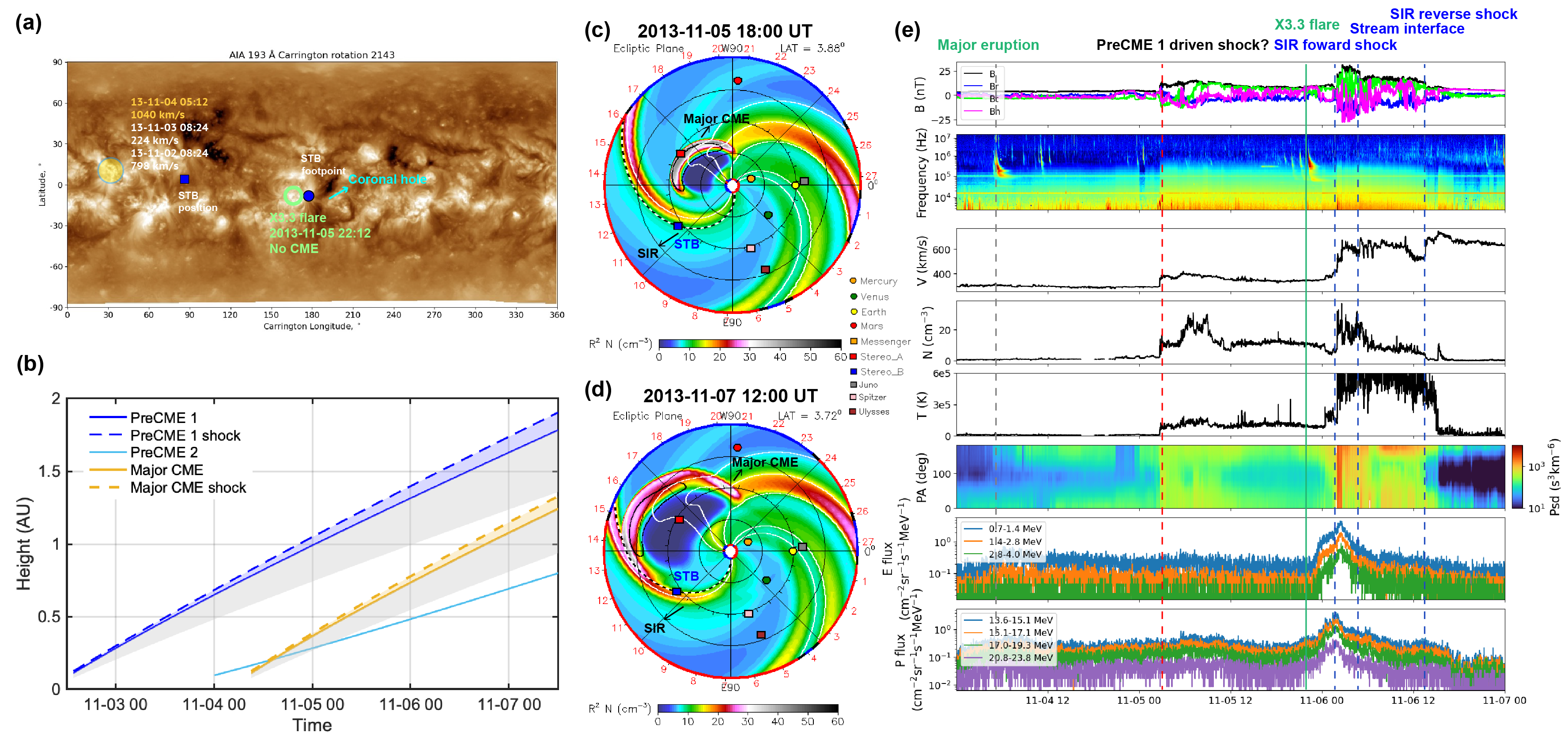}
\caption{\label{fig:cme_cir_flare}  Shock event on 2013 November 6. (a) Synoptic map showing the source regions of the major CME (yellow) and preceding CMEs (white). (b) DBM propagation results. (c)–(d) Snapshots from the WSA–ENLIL simulations. (e) in-situ observations from STA, including the magnetic field, radio dynamic spectrum, solar wind speed, proton density, temperature, PADs of 151 eV electrons, and temporal profiles of MeV electrons and protons.}
\end{figure*}

Fig. \ref{fig:cme_cir_flare} presents a case in which flare-accelerated particles may have been injected into a SIR and subsequently re-accelerated by its forward shock. To investigate the associated large-scale structure encountered by STB, we first examine the related CME activities. The fast CME closest in time to the shock occurred on 2013 November 4 at 05:12 UT, with a linear speed of 1040 km/s, and is hereafter referred to as the major CME. Two earlier CMEs originating from the same active region were also identified: PreCME 2 on November 3 at 08:24 UT with a linear speed of 224 km/s and PreCME 1 on November 2 at 08:24 UT with a speed of 798 km/s. As the DBM result in Panel (b) shows, the similar speeds of PreCME 1 and the major CME suggest little interaction between them. The WSA--ENLIL simulation from the DONKI database\footnote{\url{https://kauai.ccmc.gsfc.nasa.gov/DONKI/view/WSA-ENLIL/3564/1}}, which simulates the major CME with a half width of 70$^\circ$, however, indicates that its propagation direction is far from STB and that it is unlikely to reach the spacecraft. It suggest that an SIR structure was passing STB at that time, likely associated with the high-speed stream from the coronal hole magnetically connected to the spacecraft.

We further examine the in-situ observations shown in Panel (e). The gray dashed line marks the type III burst associated with the major eruption. As shown in Panel (a), during the major eruption, STB was magnetically connected to a coronal hole located approximately 150$^\circ$ away in longitude on the western side. No significant SEP enhancement is observed at that time, suggesting that the event was relatively weak and that the magnetic connectivity between STB and the shock was poor. As indicated by the red dashed line, about 21 h after the eruption, STB detected a weak CME-driven shock, temporally consistent with the arrival of the shock driven by the PreCME 1 predicted by DBM. Within the CME, a prolonged bi-directional streaming is observed, indicating the presence of closed magnetic field lines. At 02:00 UT on 6 November, STB observed a clear forward shock, characterized by sharp increases in the magnetic field, solar wind speed, temperature, and density. This shock was listed as a CME-driven shock by \cite{stereo_shock_list}. However, we identified a stream interface and a reverse shock, which are typical signatures of a SIR. In addition, the upstream and downstream solar wind conditions of the SIR correspond to typical slow and fast solar wind, respectively. We therefore interpret this shock as a SIR driven forward shock. As indicated by the second blue dashed line, a stream interface is identified, where the total magnetic field decreases, the solar wind speed slightly increases, and the density drops, while the temperature shows strong fluctuations without a clear trend. At the rightmost blue dashed line, a reverse shock is observed, marked by decreases in magnetic field strength, density, and temperature across the shock. This SIR appears to merge with the preceding CME, as evidenced by the presence of clear bi-directional streaming in the high-speed solar wind following the SIR, likely associated with the CME leg.

Tab. \ref{tab:shock_sir_flare} summarizes the parameters of the forward shock. Although this shock occurred in a complex SIR–CME interaction region, we suggest that the local enhancement of MeV electron and proton fluxes near the shock was likely associated with a fresh seed particle injection from the Sun. The onset of the particle increase coincides with a type III radio burst associated with an X3.3-class flare without a CME. As shown in Panel (a) of Fig. \ref{fig:cme_cir_flare}, this flare is located adjacent to the coronal hole that is magnetically connected to STB, which may provide a direct pathway for flare-accelerated particles to access the spacecraft. We therefore infer that this flare supplies the seed population for subsequent shock acceleration, consistent with the scenario briefly discussed by \cite{stereo_shock_acc}. Following a peak near the shock, the particle flux gradually decays toward background levels.

\begin{table*}
\caption{Parameters of the SIR forward shock on 2013 November 6 in Fig. \ref{fig:cme_cir_flare}.}
\label{tab:shock_sir_flare}
\centering
\renewcommand{\arraystretch}{1.3} 
\setlength{\tabcolsep}{6pt} 
\begin{tabular}{lccccccc}
\hline\hline
Date & $V_{sh}$ (km/s) & $\theta_{\mathrm{Bn}}$ ($^\circ$) & $B_d/B_u$ & $M_{fms}$ & $\beta$ & $V_A$ (km/s)  & $M_A$ \\
\hline
2013-11-06 02:00 & 549.8 $\pm$ 7.5 & 45.1 $\pm$ 2.9 & 2.3 $\pm$ 0.1 & 2.31 $\pm$ 0.04 & 1.10 $\pm$ 0.17 & 48.9 $\pm$ 3.1 & 2.92 $\pm$ 0.14  \\
\hline
\end{tabular}
\end{table*}

\section{Simulation of Electron Shock Acceleration within Upstream Magnetic Loops} \label{sec:simulation}

To investigate the geometry effect of a upstream magnetic loop on electron acceleration in comparison with an open field line scenario, we have employed the model EASI \citep{nyberg2026}. EASI is an open-source\footnote{\url{https://github.com/SeveNyberg/easi}} one-dimensional (1D) Monte Carlo test particle model, where particles are traced under the guiding-center approximation, focused in magnetic field gradients, and stochastically scattered using an ad hoc mean free path profile. The simulation takes a set of physical input parameters, namely the shock speed, Alfv\'en speed, shock obliquity, upstream plasma density, and upstream plasma temperature. The shock has a finite thickness (of the order of an ion inertial length $d_{\rm i}$) and uses a hyperbolic tangent profile for the transition from the upstream to the downstream. For more details on the model, see \cite{nyberg2026}. 

\begin{figure}
\includegraphics[width=1.0\linewidth]{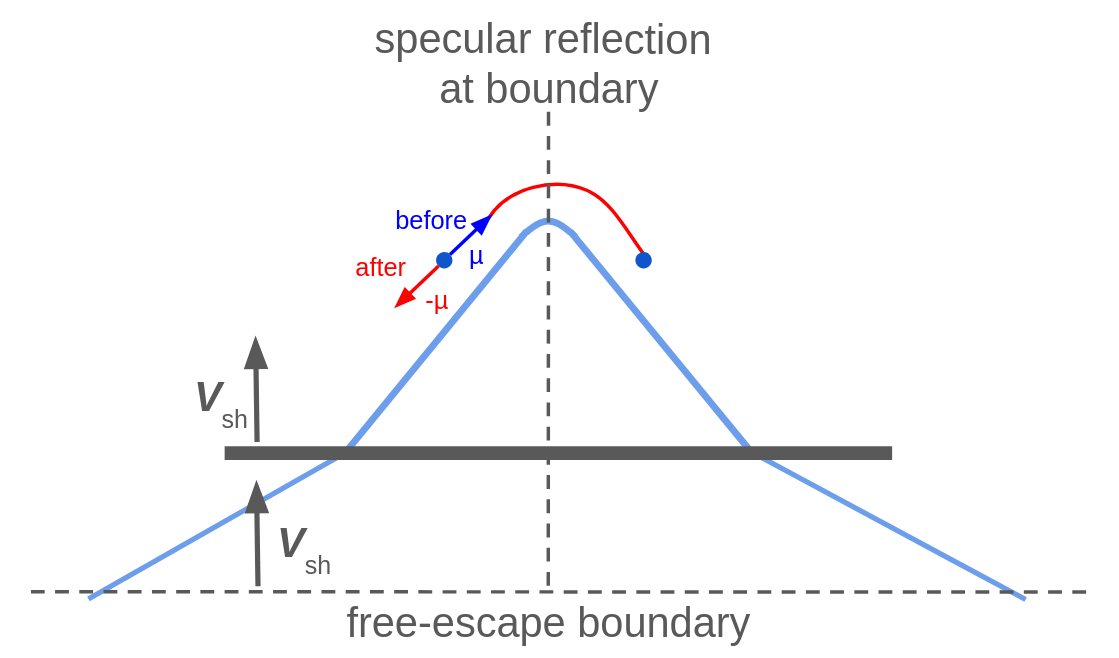}
\caption{The simulation geometry in the simulation case with a magnetic loop. A specular reflection boundary, where a particle crossing the boundary will be returned to the box with its (fluid-frame) pitch angle cosine $\mu$ flipped to $-\mu$, is set before the turn in the magnetic loop to emulate particles symmetrically traversing the loop legs. The upstream simulation box size decreases as the shock traverses towards the magnetic loop apex.}
\label{fig:easi_geom}
\end{figure}

The simulation has a finite box size, with the upstream having either a free-escape boundary in the case of the open field line or a specular reflection boundary in the case of the magnetic loop as presented in Fig. \ref{fig:easi_geom}. This set-up corresponds to a symmetric system where the field line is closed and the shock is propagating towards a loop apex (or cusp) along both legs of the loop with equal strength at both legs. The downstream boundary is a free-escape boundary in both cases. The size of the upstream in the case of the open field line and the downstream box size is set to the diffusion length corresponding to injection energy of the particles, 
$L_{\rm diff}(v_{\rm inj}) = \frac{1}{3}\cdot \lambda_1 \cdot \cos^2 \theta_1 \cdot v_{\rm inj}/u_{x,1}$,
where $\lambda_1$ is the upstream ambient mean free path of particles, $\theta_1$ is the shock obliquity, $v_{\rm inj}$ is the injection speed of particles, and $u_{x,1}$ is the upstream plasma flow speed along the shock normal in the shock rest frame. For the magnetic loop case, the upstream box size along the shock normal direction starts from a set value of 5 $R_\odot$. This value is based on the estimated size of the sub-flux rope immediately upstream of the shock observed by STA on 2012 January 29 (see Sec. \ref{sec:shock_in_cme} and Fig. \ref{fig:shock_mfr}). To simulate the propagation of the shock through the magnetic loop we consecutively reduce the box size through the simulation run until the upstream boundary crosses the shock to the downstream side ($x=0$), where the simulation then ends. The maximum in-simulation time used in the case of the open field line corresponds to the time the shock front takes to propagate to the magnetic loop apex, $t_{\rm max} = L_{\rm loop} / u_{x,1}$, where $L_{\rm loop} = 5\ R_\odot$.

\begin{figure}
\includegraphics[width=\linewidth]{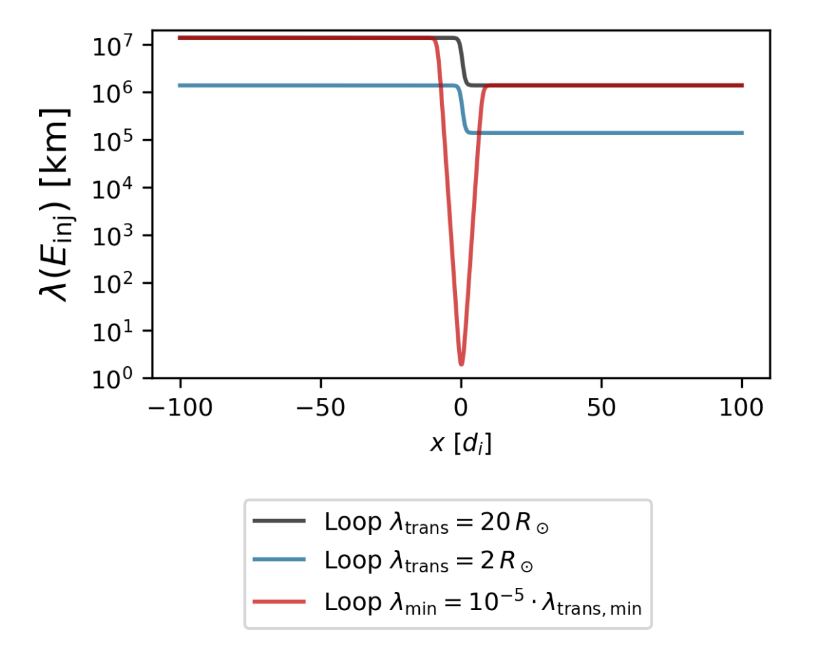}
\caption{The different mean free path $\lambda$ profiles used in the simulations. The values indicated by $\lambda_{\rm trans}$ are upstream values of the mean free path, with a tenfold reduction to the downstream. The reduction at the shock contains a multiplier that adjusts the bottom of the reduction in comparison to the downstream mean free path value.}
\label{fig:easi_prof}
\end{figure}

To investigate particle acceleration, different ad hoc mean free path profiles were chosen. Following \cite{nyberg2026}, a hyperbolic tangent transition profile through the shock and a hyperbolic cosine modulated reduction at the shock transition were chosen (Fig. \ref{fig:easi_prof}). The highest mean free path case, 20 $R_\odot$ with no reduction in the transition region, corresponds to a nominal solar wind mean free path with a classic understanding of the shock transition. The smaller mean free path, 2 $R_\odot$ with no reduction at the shock, was chosen as a comparison case to see how a further reduction of the mean free path will affect particle acceleration. As shown in Fig.~\ref{fig:wave}, high-frequency wave activity is observed during the $\sim$8 s immediately upstream of the shock. Motivated by this observation, we also consider a localized reduction of the mean free path confined to the shock transition (red curve in Fig.~\ref{fig:easi_prof}). The reduced mean free path cases correspond to the SSDA theory \citep{katou2019,amano2020}, where the shock transition region traps particles with turbulence generated by, e.g., current-driven instabilities.

\begin{figure}
\includegraphics[width=\linewidth]{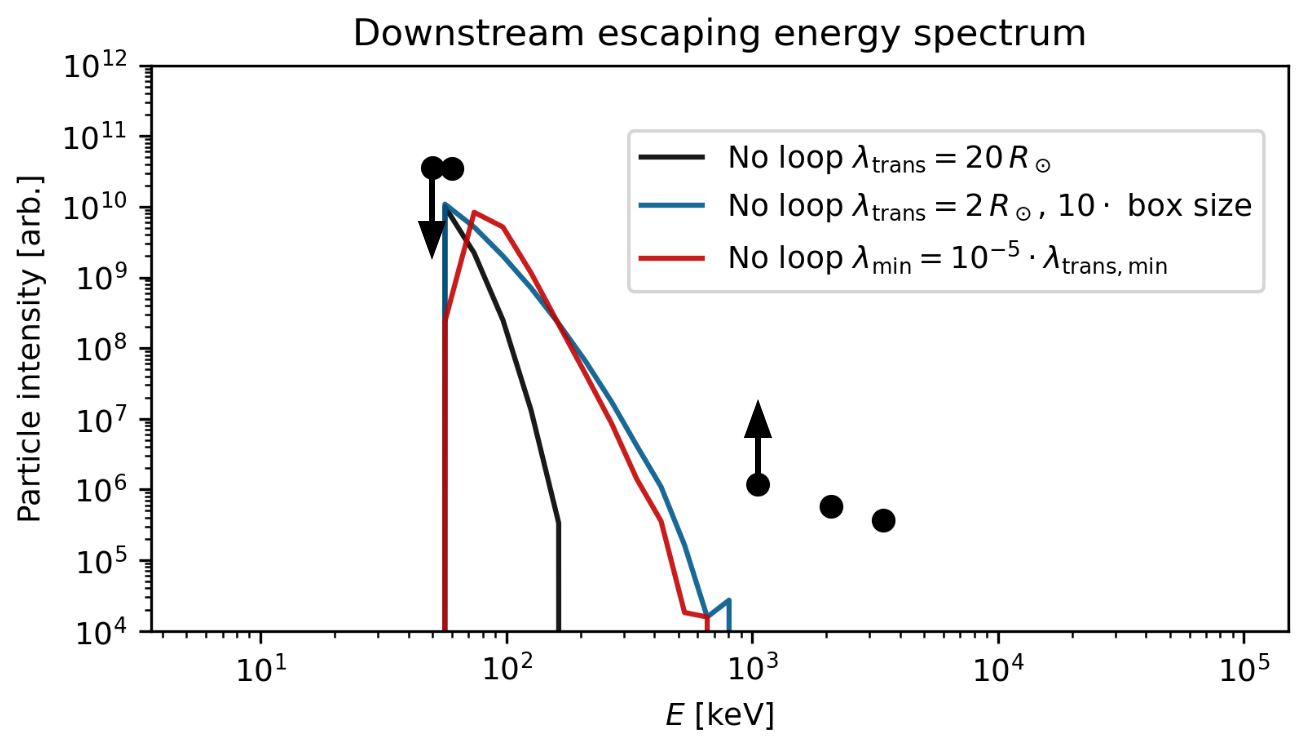}
\caption{Particle intensity spectra from the simulation with an open field-line geometry, compared with STA observations. Only the lowest energy channels of SEPT ($\sim$60 keV) are shown, as the higher-energy channels are strongly contaminated by ions. The arrows associated with the SEPT and HET measurements indicate the possible range arising from ion contamination and inter-calibration, respectively.}
\label{fig:easi_open}
\end{figure}

\begin{figure}
\includegraphics[width=\linewidth]{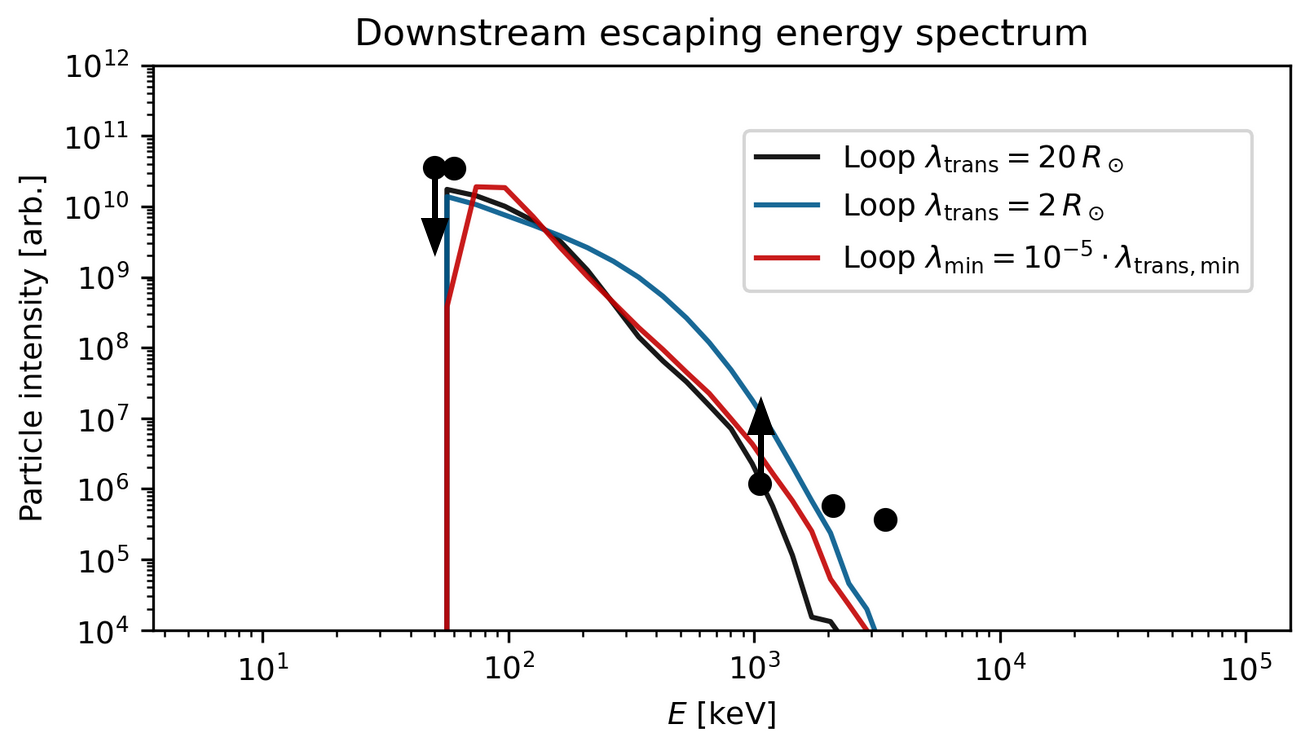}
\caption{The particle intensity spectra for the results of the simulation with a magnetic loop geometry and observations by STA as in Fig.~\ref{fig:easi_open}. }
\label{fig:easi_loop}
\end{figure}

Based on the observed parameters of the 2012 January 29 shock in Tab. \ref{tab:shock}, an upstream flow speed $u_{x,1}$ of 175 km/s, an Alfv\'en speed of 55 km/s, a shock obliquity of 86.8$^\circ$, a temperature of 36650 K, and a particle density of 17.72 cm$^{-3}$ were used as the physical input parameters of the simulation. The particles were injected monoenergetically at 50 keV. This assumes the presence of seed electrons, consistent with the observed energetic electron population trapped in the preceding CME upstream of the shock. A maximum simulation time of 19886 s ($\sim$5.5 h) was used for the open field-line geometry. The resulting particle intensity spectra for the open field line and magnetic loop cases can be seen in Figs. \ref{fig:easi_open} and \ref{fig:easi_loop}, respectively. The simulation spectra are created from Monte Carlo particles in the downstream side of the shock at the end of the simulation, as in the magnetic loop case, no particles are left in the upstream as the upstream box size reduces to zero. To facilitate comparison with the simulation results, the observed peak spectrum at the shock front is also shown together with the SEPT and HET measurements. Because the SEPT electron data are strongly contaminated by ions, only the lowest-energy channels ($\sim$60 keV) are used, with the arrows indicating the direction of the margin of error due to ion contamination. For the HET measurements, the first energy channel ($\sim$1 MeV) shown in the spectrum has been multiplied by a factor of 10, following the inter-calibration factor derived for STA/HET measurements \citep{2025A&A...693A.198F}. 

In the open field-line geometry, adopting a globally reduced mean free path ($\lambda_{\rm trans}=2\,R_\odot$) or a reduction confined to the shock transition leads to substantially more efficient electron acceleration than the reference case ($\lambda_{\rm trans}=20\,R_\odot$). Both reduced-mean-free-path cases reach similar maximum energies, indicating that efficient acceleration is primarily controlled by enhanced scattering near the shock, which prolongs particle residence time. The globally reduced case yields a smoother broken power-law spectrum, whereas the locally reduced case preserves greater spectral complexity.

In the magnetic-loop geometry, particle acceleration is considerably more efficient, and all three mean free path cases accelerate electrons to energies above 1 MeV. Compared with the reference case ($\lambda_{\rm trans}=20\,R_\odot$), reducing the mean free path further increases the maximum energy, although the improvement is less pronounced than in the open field-line geometry. The spectra from the reference case and the locally reduced case retain more spectral complexity, whereas the globally reduced case produces a smoother spectrum. Overall, the magnetic-loop geometry reproduces the observed spectral shape more closely.

\section{Discussion and Conclusions}\label{sec:conclusion}

In this study, we investigated nine interplanetary shocks associated with MeV electron acceleration from their solar origins to in-situ observations, focusing on three scenarios that may enhance shock acceleration: shocks propagating within preceding CMEs, shock–SIR interactions, and re-acceleration of newly injected flare-accelerated electrons. Fig. \ref{fig:pad} presents the in-situ observations within $\pm$2 h of all nine events. Six events correspond to shocks propagating within preceding CMEs, indicated by the blue boxes marking the shock times in the figure titles. Among the remaining three events, the 2011 June 5 event, discussed in Sec. \ref{sec:SIR}, involved multiple preceding CMEs and the shock–SIR interaction, resulting in a highly complex shock downstream. The 2012 July 23 event was a historically strong shock and associated with a twin-CME eruption \citep{Russell_super_shock,Liu_super_shock}. For the 2017 July 24 event, we identified another ICME-driven shock approximately 70 min after the shock associated with the MeV electron enhancement. The two shocks are associated with a CME sequence from the same active region, consisting of a fast CME and two preceding intermediate-speed CMEs that erupted about 10 min and 3 h before the fast CME. Overall, preceding CMEs are a common feature among these nine events and may play an important role in shaping the conditions for shock acceleration by influencing the shock geometry, large-scale particle trapping, and the seed population.

For the scenario of a shock in a preceding CME discussed in Sec. \ref{sec:shock_in_cme}, we analyzed two representative events in detail, on 2012 January 29 and 2012 May 28. The two shocks propagated through a large-scale CME and a smaller magnetic flux rope, respectively. Both were low-$\beta$, quasi-perpendicular shocks, consistent with the statistical results on shocks inside ICMEs by \cite{shock_in_icme}. However, the two shocks represent different conditions within this population. The 2012 January 29 event was an extreme case, with an upstream magnetic-field magnitude of $\sim$18 nT, approaching the highest values among the 49 shocks examined by \cite{shock_in_icme}. In contrast, the upstream magnetic field of the 2012 May 28 shock was $\sim$9 nT, closer to the typical magnetic-field strength inside ICMEs at 1 au \citep{richardson2010near}. Besides, the two events highlight different processes that may influence particle acceleration. The 2012 January 29 event was characterized by a relatively steady shock and stable upstream and downstream conditions. The strong magnetic field within the upstream flux rope likely provided large-scale magnetic confinement, making this event an illustrative example of how an organized upstream structure can influence particle transport and trapping around a shock. In contrast, the 2012 May 28 event exhibited a pronounced magnetic overshoot at the shock ramp, suggesting that the local shock structure itself may have played an important role in particle reflection and acceleration.

We also investigated the roles of shock--SIR interactions and re-acceleration of flare-related seed particles in the 2011 June 5 and 2013 November 6 events, respectively. While the effects of SIRs on energetic-particle transport and their interactions with CMEs have been extensively studied \citep[e.g.,][]{Lario_SIR,Nicolas_SIR}, their role in shaping the seed-particle population has also been demonstrated for ions. For example, \cite{2023JGRA..12831203W} showed that corotating interaction regions (CIRs) and merged interaction regions can substantially modify the spatial and energy distributions of suprathermal ions, with localized enhancements potentially serving as seed populations for subsequent CME-driven shocks. Whether SIRs produce similar modifications to the electron seed population remains less clear. Recent observations of electron acceleration at CIR shocks have suggested that the acceleration may be primarily governed by SDA \citep{guo2024evolution}. Our results further suggest that, with seed particles supplied by an energetic solar eruption, the SIR shock pair may further re-accelerate these electrons. During the prolonged interaction between the ICME and the SIR, the particle acceleration and transport can be highly complex. For the 2013 November 6 event, the X-class flare was magnetically connected to the SIR, providing a fresh population of energetic seed particles directly to the SIR forward shock. This event therefore indicates the importance of magnetic connectivity and the availability of seed particles in enabling efficient particle acceleration at an interplanetary shock.

As a representative shock-in-ICME event, the shock on 2012 January 29 was investigated in detail, and electron acceleration was simulated based on the observed shock and upstream conditions. As presented in Sec. \ref{sec:shock_in_cme}, the shock encountered a $\sim$2 h sub-flux-rope structure within the preceding CME prior to its arrival. Suprathermal electrons at both 151 and 1075 eV exhibited clear bidirectional streaming, indicating closed magnetic field lines upstream of the shock. These observations provide the basis for the closed upstream magnetic-loop geometry adopted in our simulations, in which particles undergo specular reflection at the loop. Such reflection by preceding ICMEs has also been observationally reported for energetic electrons, producing counter-streaming electron beams and, in some cases, inverse velocity dispersion signatures \citep{Liuzzo_reflectron}. The absence of similarly clear bidirectional signatures in most of the other events shown in Fig. \ref{fig:pad} does not necessarily imply the absence of particle trapping, as the observed beam signatures can depend sensitively on the magnetic geometry and pitch-angle scattering \citep{flossie2026modelling}. In this case, the nearly perpendicular shock geometry ($\theta_{Bn}=87^\circ$) further provides a suitable configuration for investigating electron acceleration at a quasi-perpendicular shock.

As shown in Fig.~\ref{fig:wave}, high-frequency ($\sim$10 Hz) magnetic fluctuations are observed during the $\sim$8 s immediately upstream of the 2012 January 29 shock. In a statistical study, \cite{wilson2017revisiting} found that upstream whistler precursors are commonly observed ahead of low-$\beta$, quasi-perpendicular shocks. The 2012 January 29 shock exhibits similar conditions, with $\beta_i\sim0.08$ and $\theta_{Bn}\sim87^\circ$. The relative fluctuation amplitude ($\delta B/B\sim0.06$ for $\tau=0.1$ s) is lower than the typical values reported by \cite{wilson2017revisiting}, and the fluctuations persist for only a short interval. Moreover, because our wave analysis is limited to frequencies below 16 Hz, corresponding to less than $\sim0.1f_{ce}$ for this event, the full properties of these high-frequency fluctuations cannot be resolved. Nevertheless, their presence immediately upstream of the shock motivates our consideration of enhanced scattering localized at the shock transition in the simulations.

Our simulations also show that, in the open field-line geometry, enhanced scattering confined to the shock transition produces nearly the same maximum energies as the globally reduced mean free path, suggesting that scattering near the shock is sufficient to enhance electron acceleration. In contrast, the magnetic-loop geometry is much less sensitive to the assumed mean free path profile, with all three mean free path cases producing similar maximum energies (Fig.~\ref{fig:easi_loop}), indicating that large-scale magnetic trapping plays the dominant role. These results highlight the complementary roles of localized scattering and large-scale upstream magnetic topology in electron acceleration. Further investigation of field line geometry induced effects on particle transport and wave generation self-consistently should be conducted to understand the complex nature of electron acceleration in shocks. Furthermore, investigation of the effects of the seed particle population on the wave population in charge of scattering the particles, such as the investigation done on ion acceleration by \cite{nyberg2024}, could provide further understanding of energization of electrons in shocks in the context of complex solar events. We also draw attention to the fact that the model we have applied in the present study only accounts for trapping in loop-like structures but omits the potentially important effects of temporally changing shock geometry \citep[see, e.g.,][]{2006Sandroos}.

Our results indicates the importance of the upstream magnetic geometry established by preceding CMEs for electron acceleration. Compared to protons, electrons have much smaller gyroradii and travel rapidly along the magnetic field line. Their acceleration are therefore more strongly constrained by magnetic connectivity. Previous twin-CME studies have mainly attributed enhanced SEP acceleration to increased upstream turbulence and enhanced seed populations \citep{li2012twin}. Recent observations and simulations have further shown that CME interactions can involve multiple particle acceleration processes \citep[e.g.,][]{2024ApJ...967L..35N,2024ApJ...967L..33C}. While enhanced turbulence is crucial to diffusive shock acceleration, electron acceleration at quasi-perpendicular shocks can be strongly affected by the upstream magnetic geometry through SDA. The complex magnetic environment created by preceding CMEs may introduce substantial spatial variations in magnetic connectivity and shock geometry, potentially allowing different acceleration and transport processes to operate in different regions of the same shock. 

A recent model of electron acceleration at Earth’s bow shock proposed that electrons can be trapped between the bow shock and a foreshock-transient boundary. This trapping enables repeated interactions with the shock and the moving magnetic boundary, while resonant wave scattering, adiabatic reflection, and SDA jointly contribute to electron acceleration \citep{shi2025compound}. Our model considers a relatively simple magnetic trapping configuration, whereas interplanetary shocks propagating through large-scale magnetic transients may encounter more complex shock and foreshock structures that are not examined in detail here. Further in-depth studies of these complex events may therefore help to figure out the role of magnetic trapping and shock structures in electron acceleration under complex upstream conditions.

\begin{acknowledgements}

We thank the anonymous referee for the valuable comments and suggestions. We acknowledge the use of data from the STEREO, SDO and SOHO spacecraft. We also acknowledge the Community Coordinated Modeling Center (CCMC) at Goddard Space Flight Center for providing the WSA--ENLIL+CONE model results through the DONKI database at \url{https://ccmc.gsfc.nasa.gov/tools/DONKI/}. Python wavelet software was provided by Evgeniya Predybaylo, based on \cite{torrence1998practical}, and is available at \url{http://atoc.colorado.edu/research/wavelets/}. X.C. and C.L. acknowledge the support by NSFC under grant 12333009 and by the Fundamental Research Funds for the Central Universities under grant KG202506. I.C.J. was supported by the Research Council of Finland (X-Scale, grant No.~371569) and by FWO grant No.~1295826N. Work in the University of Turku was performed under the umbrella of Finnish Centre of Excellence in Research of Sustainable Space (FORESAIL) and Space Resilience funded by the Research Council of Finland (grant No.\ 352847 and No.\ 374097). N.D. acknowledges support by the Research Council of Finland (SHOCKSEE, grant No.\ 346902 and AIPAD, grant No.\ 368509).

\end{acknowledgements}

\bibliographystyle{aa} 
\bibliography{ref} 

\appendix

\section{Source region of the nine shock events} \label{sec:sources}

Figure~\ref{fig:source1} presents the solar source analysis for all nine shock events investigated in this study, extending the representative example shown in Fig.~\ref{fig:source}. The same analysis procedure and plotting conventions described in Sec.~\ref{sec:source_propagation} are applied throughout. 

\begin{figure*}[htb!]
    \begin{center}
    \includegraphics[width=0.9\textwidth]{euv_synoptic_map_all.png}
    \caption{Source regions of the associated CMEs in the nine shock events. The top left panel shows the timing relationship between the major CMEs and their preceding CMEs for all events. The last nine panels present the source regions of the major and preceding CMEs with yellow circles. The blue lines indicate the great-circle distances between the two active regions on the solar surface, while the red (blue) dots and squares mark the footpoints and line-of-sight positions of STA(STB). \label{fig:source1}}
    \end{center}
\end{figure*}

\section{Zoom-in observation of shocks}

\begin{figure*}[htb!]
\centering
\includegraphics[width=1\linewidth]{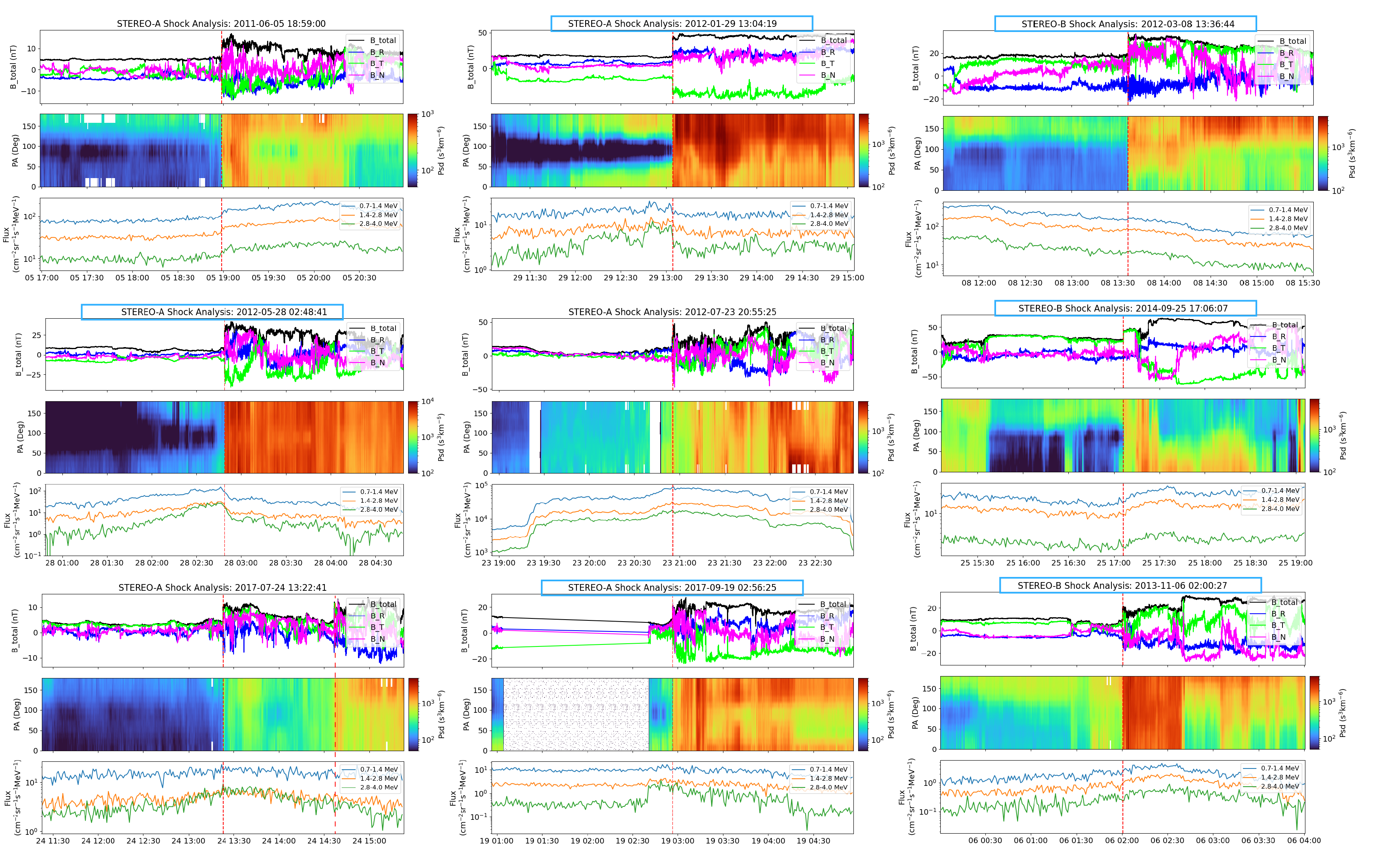}
\caption{\label{fig:pad} Zoom-in ($\pm$2 h) observations around the shocks. From top to bottom, the panels show the magnetic field, electron PADs at 151 eV, and HET electron intensity profiles.}
\end{figure*}

Figure~\ref{fig:pad} presents zoom-in ($\pm$2 h) observations of the magnetic field, 151 eV electron PADs, and MeV electron intensity profiles around the shocks for all nine events investigated in this study.

\section{Wave analysis of 2012 January 29 shock event}

\begin{figure*}[htb!]
\centering
\includegraphics[width=1\linewidth]{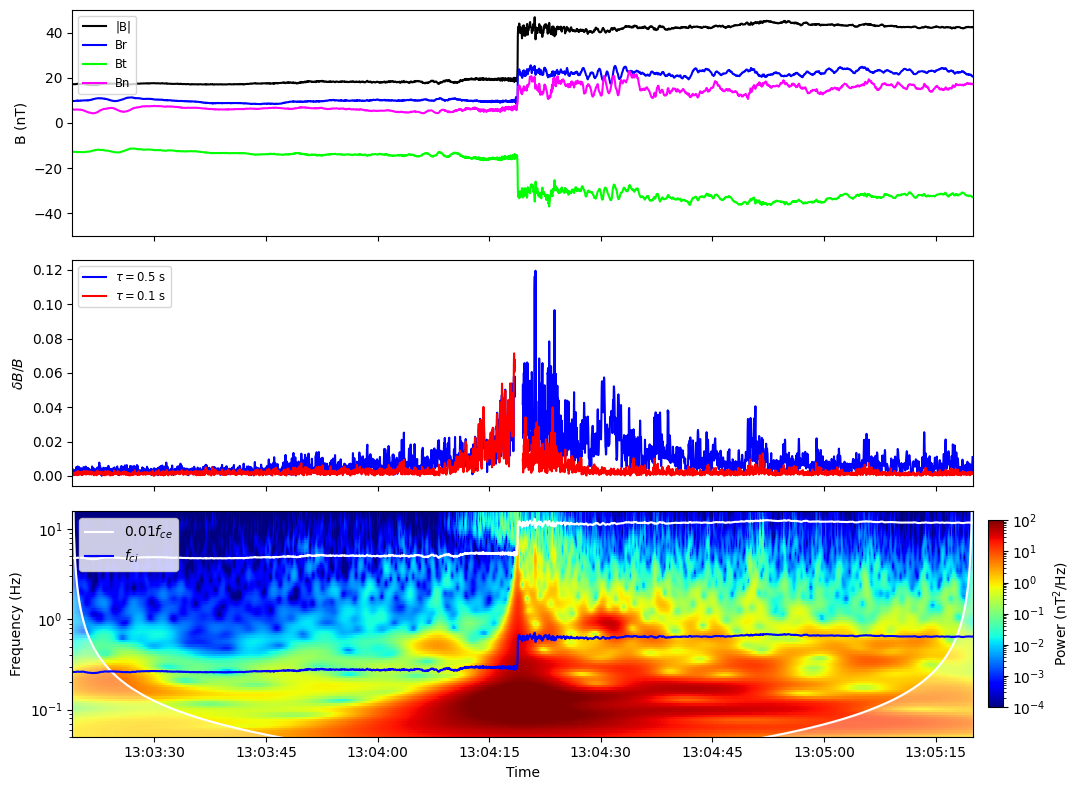}
\caption{\label{fig:wave}Wave observations of the 2012 January 29 shock. From top to bottom, the panels show the magnetic field measured in the 32 s$^{-1}$ burst mode, the magnetic field fluctuation amplitudes at $\tau=0.1$ s and $\tau=0.5$ s, and the corresponding Morlet wavelet power spectrum.}
\end{figure*}

Fig.~\ref{fig:wave} presents the upstream wave observations for the 2012 January 29 shock event. It includes the 32 s$^{-1}$ magnetic field measurements, normalized magnetic field fluctuations, and the corresponding Morlet wavelet power spectrum.

The wave analysis was performed using the 32 s$^{-1}$ burst-mode magnetic field data. To characterize high-frequency magnetic fluctuations, we calculated the normalized fluctuation amplitude,
\begin{equation}
\delta B = \left|\mathbf{B}-\langle\mathbf{B}\rangle\right|,
\end{equation}
where $\langle\mathbf{B}\rangle$ is the running mean of the magnetic field vector over a timescale $\tau$. The normalized fluctuation amplitude is then defined as
\begin{equation}
\frac{\delta B}{B}
=
\frac{\left|\mathbf{B}-\langle\mathbf{B}\rangle\right|}
{\left\langle |\mathbf{B}| \right\rangle},
\end{equation}
where $\left\langle |\mathbf{B}| \right\rangle$ is the running mean of the magnetic field magnitude. Since energetic electrons are expected to interact primarily with relatively high-frequency fluctuations, we focus on two characteristic timescales, $\tau=0.1$ s and $\tau=0.5$ s. To emphasize the upstream magnetic fluctuations, the values of $\delta B/B$ within the shock ramp ($\pm0.5$ s around the shock crossing) were masked in Fig.~\ref{fig:wave}. 

The magnetic field wavelet power spectrum was derived using the Morlet wavelet analysis following \cite{torrence1998practical}. The local proton cyclotron frequency $f_{ci}$ and 0.01 times the electron cyclotron frequency $0.01\times f_{ce}$ were overplotted on the wavelet spectrum for reference.

\end{document}